# The FMRIB Variational Bayesian Inference Tutorial II:
# Stochastic Variational Bayes

Michael A. Chappell[1,2,3] & Mark W. Woolrich[4]

michael.chappell@nottingham.ox.ac.uk

[1] Sir Peter Mansfield Imaging Centre, School of Medicine, University of Nottingham, UK.
[2] Institute of Biomedical Engineering, Department of Engineering Science, University of Oxford, UK.
[3] Wellcome Centre for Integrative Neuroimaging, FMRIB, Nuffield Department of Neurosciences, University of Oxford, UK.
[4] Oxford Centre for Human Brain Activity, Wellcome Centre for Integrative Neuroimaging, Department of Psychiatry, University of Oxford, UK.

Example code and exercises associated with this tutorial can be found here:
https://vb-tutorial.readthedocs.io

This work is licensed under a Creative Commons Attribution 4.0 International License (CC-BY 4.0).
This work was supported in part by the Engineering and Physical Sciences Research Council UK (EP/P012361/1). The Wellcome Centre for Integrative Neuroimaging is supported by core funding from the Wellcome Trust (203139/Z/16/Z). MWW's research is supported by the NIHR Oxford Health Biomedical Research Centre, by the Wellcome Trust (106183/Z/14/Z), and the MRC UK MEG Partnership Grant (MR/K005464/1).

# Introduction

Bayesian methods have proved powerful in many applications, including MRI, for the inference of model parameters from data, e.g. the use of physiological models to interpret functional MRI time-series data. These methods are based on Bayes' theorem, which itself is deceptively simple. However, in practice the computations required are intractable even for simple cases. Hence methods for Bayesian inference have historically either been significantly approximate, e.g., the Laplace approximation, or achieve samples from the exact solution at significant computational expense, e.g., Markov Chain Monte Carlo methods. Since around the year 2000 so-called Variational approaches to Bayesian inference have been increasingly deployed, in many cases based on the method proposed by (Attias et al. 2000). In its most general form Variational Bayes (VB) involves approximating the true posterior probability distribution via another more 'manageable' distribution, the aim being to achieve as good an approximation as possible. In the original FMRIB Variational Bayes tutorial (Chappell and Woolrich 2016) we documented an approach to VB based on (Attias et al. 2000) that took a 'mean field' approach to forming the approximate posterior, required the conjugacy of prior and likelihood, and exploited the Calculus of Variations, to derive an iterative series of update equations, akin to Expectation Maximisation, for Bayesian inference. In this tutorial we revisit VB, but now take a stochastic approach to the problem that potentially circumvents some of the limitations imposed by the earlier methodology. This new approach bears a lot of similarity to, and has benefited from, computational methods applied to machine learning algorithms, particularly (Kingma and Welling 2013). Although, what we document here is still recognisably Bayesian inference in the classic sense, and not an attempt to use machine learning as a black-box to solve the inference problem.

# Bayesian Inference

The basic Bayesian inference problem is one where we have a series of measurements, *y*, and we wish to use them to determine the parameters, *w*, of our chosen model $M$. The method is based on Bayes' theorem:

$$p(w|y, M) = \frac{p(y, w|M)}{p(y|M)} = \frac{p(y|w, M)p(w|M)}{p(y|M)} \qquad (1.)$$

Which gives the *posterior* probability of the parameters given the data and the model, $p(w|y, M)$, in terms of: the *likelihood* of the data given the model with parameters *w*, $p(y|w, M)$, the *prior* probability of the parameters for this model, $p(w|M)$, and the *evidence* for the measurements given the chosen model, $p(y|M)$. If we are not too concerned with the correct normalisation of the posterior probability distribution, we can neglect the evidence term to give:

$$p(w|y) \propto p(y|w)p(w) \qquad (2.)$$

Where the dependence on the model, *M*, is implicitly assumed.

From the definition of Bayes theory in equation (2.) we can start to describe a practical method for Bayesian inference: write down the likelihood for the data in question based on the model, *M*, that should be a full description of the data generating process including noise; write down a prior distribution capturing a prior knowledge about the model parameters; multiply the two and hence form the posterior distribution (up-to-scale). In practice, what we will want from this at least is a best estimate of the parameter(s), *w*, from the posterior distribution, if not also some measure of uncertainty. A suitable statistical approach would be to take moments of the distribution, e.g., the mean and the variance. To do that requires us perform an integral on our posterior distribution, and this is generally where we have a problem: the posterior distribution is, in most cases, intractable.

> Note that we cannot easily get around this problem. If instead we were to try to take confidence intervals of the distribution rather than calculate the variance, we would need to correctly scale the posterior distribution. This would itself involve an integration. The problem is exactly the same if we go back to the version of Bayes' theorem in Equation (1.). This requires us to calculate the evidence term (a useful term in its own right), but to do so requires integration. The best we can do with the posterior only up-to-scale is to find the maximum, i.e., the mode of the distribution – so-called Maximum A Posteriori inference – and use a measure of the local curvature of the distribution to say something about uncertainty (in essence this is the Laplace Approximation).

## Variational Inference

If our posterior distribution happened to be from one of the small group of 'known' distributions, generally ones we can calculate moments of, we would be okay. Variational Bayes, in its most general sense, involves taking a known distribution as an *approximate posterior*, $q(\theta)$, and trying to find the version of it that is as close as possible to the true posterior. To measure 'closeness' we use the Kullback-Liebler (KL) divergence (or distance) between the two distributions. In practice, this is something we cannot calculate because it requires integration of the posterior, but minimising the KL divergence between the approximate posterior and the true posterior is equivalent to maximizing the free energy[1]:

$$F(\theta) = \int q(\theta) \log \left( p(y|\theta) \frac{p(\theta)}{q(\theta)} \right) d\theta \tag{3.}$$

This is something that we describe in more detail in the original FMRIB VB tutorial (Chappell and Woolrich 2016). Box 1 gives some added insight into how we might interpret the form of this Free Energy expression.

One option to maximise *F* is to choose a parameterised form for the approximate posterior: $q(\theta; \zeta)$, where $\zeta$ is the set of hyper-parameters. We can then appeal to the Calculus of Variations to arrive at an expression under which we can derive a series of 'update equations' for the set of hyper-parameters of the approximate posterior. Since this involves an integration, this method places a number of constrains on the choice of the approximate posterior distribution. For example, the use of conjugate priors. This was the subject of the original FMRIB Variational Bayes Tutorial(Chappell and Woolrich 2016) and what is notable about this approach, quite apart from the constraints, is the amount of 'manual' integration required that is specific to the likelihood/posterior in question and thus needs to be repeated if there are any changes to the model.

---

[1] The free energy is a lower bound on the model evidence and thus is alternatively called the 'Evidence Lower BOund' or ELBO.

> **Box 1: Interpreting the form of the free energy equation**
> We can recognise the form of F as being an expectation over $q(\theta)$:
> $$F = E_{q(\theta)}\left[\log\left(p(y|\theta)\frac{p(\theta)}{q(\theta)}\right)\right]$$
> We can split the terms and write the free energy as:
> $$F = E_{q(\theta)}[\log p(y|\theta)] - E_{q(\theta)}\left[\log\left(\frac{q(\theta)}{p(\theta)}\right)\right]$$
> This illustrates that the process of maximizing the free energy is a combination of maximizing the log-likelihood and minimizing the KL divergence between the (approximate) posterior and the prior. The first term encourages parameter values that explain the observed data, the second term favours posterior distributions that are close to the prior.
>
> We might equivalently write F in terms of the joint probability as:
> $$F = \int q(\theta) \log\left(\frac{p(y,\theta)}{q(\theta)}\right) d\theta$$
> $$= E_{q(\theta)}[\log p(y,\theta)] - E_{q(\theta)}[\log(q(\theta))]$$
> Under this formulation the first term represents and energy and encourages $q(\theta)$ to focus probability mass where the model puts high probability, $p(y,\theta)$. The second term (including the minus sign) is the entropy of $q(\theta)$ and encourages $q(\theta)$ to spread probability mass to avoid concentrating it in one location.

## Stochastic Variational Bayes

An alternative to the analytical approach considered above would be to attempt a 'brute force' approach and attempt to maximise *F* directly using *Gradient Descent*, this will require us to be able to compute the gradients of *F* with respect to the hyper-parameters, $\zeta$, of the approximating posterior:

$$\nabla_\zeta F(\theta) = \nabla_\zeta \left(\int q(\theta) \log\left(p(y|\theta)\frac{p(\theta)}{q(\theta)}\right) d\theta\right) \quad (4.)$$

The first problem with this approach is that to compute the free energy we need to compute an integral and chances are that for the problem we are interested in (i.e., the particular likelihood, prior and posterior distribution combination) this will not be tractable.

We can potentially get around this issue by sampling, i.e., taking a Monte Carlo approximation to the integral:

$$F \approx \frac{1}{L}\sum_L \log\left(p(y|\theta^{*l})\right) - \log\left(\frac{q(\theta^{*l})}{p(\theta^{*l})}\right) \quad (5.)$$

Where $\theta^{*l}$ are drawn from $q(\theta)$. Thus, we can write the gradient as:

$$\nabla_\zeta F \approx \frac{1}{L}\sum_L \nabla_\zeta \left(\log\left(p(y|\theta^{*l})\right) - \log\left(\frac{q(\theta^{*l})}{p(\theta^{*l})}\right)\right) \quad (6.)$$

Box 2 provides some further insight into why this formulation, using sampling, does provide the approximation we need.

> **Box 2: Monte Carlo approximation to an expectation**
> To use the approximations in equations (5) and (6) we draw L samples from the approximate posterior evaluate the expression $\log(p(y|\theta^{*l}) - \log\left(\frac{q(\theta^{*l})}{p(\theta^{*l})}\right)$ for each one and then take the average. Note that this is not a classic numerical approximation to the the integral (which would itself be a challenge as the limits run from minus to plus infinity); but, exploits the fact that this expression is an expectation (see Box [X]), and that we can numerically approximate an expectation over a distribution by sampling from the distribution and computing the expression we want the expectation of and summing. This is an example of 'importance sample' for Monte Carlo integration.

This step alone means we have gone from doing Gradient Descent to now using *Stochastic Gradient Descent*, since we are reliant on samples and hence we will not get an identical result for the approximate gradient of F every time we calculate it. This leaves us with a couple of issues that we will need to consider in practice: how large should *L* be, i.e., how many samples is sufficient; and, related to this issue, how well will our Stochastic Gradient Descent converge? The latter will be an important question since any Gradient Descent scheme is iterative and relies on taking a step in parameter space along the direction indicated by the gradient, but in our implementation at each step we are computing a stochastic approximation to that gradient. At the very least, even if we start from the same initial point every time, we are unlikely to take the same path through parameter space to the maximum value as we would under a traditional Gradient Descent.

One issue that remains for us is to find an efficient way to compute the gradients required. If we were to attempt to generate an analytic expression for the gradient we would find ourselves using the chain rule to handle the various operations involved. Helpfully, the idea of assembling the gradient of a complex function via a series of individual differentiations of sub-functions via the chain rule can be generalised in the form of Automatic Differentiation. Broadly, AD is a process in which each operation in the complex function is associated with its differential. When a calculation is performed, by performing all of the relevant operations, the differential is also computed by combing the differentials of the functions via the chain rule. Conveniently, such methodology is now widely implemented in machine learning in the form of 'back-propagation', where it is used to compute the gradient of an objective function that itself is composed of a network of nodes[2]. To exploit back-propagation in this context all we need to see is that our objective function can be written in terms of a network (see Box 3). The result of all of this is that we can compute the gradient 'on-the-fly'. As long as we can write our objective function in terms of sub-functions and operations that are amenable to AD/back-propagation (and even more ideally, in terms of functions that are present within our chosen computational library), we never need to manually do any differentiation, but simply write down the objective function itself and then request the relevant gradients to be computed.

---

[2] For completeness, formally 'back-propagation' is AD in reverse accumulation mode.

> **Box 3: AD, back-propagation and functions as networks**
> Many machine learning algorithms are based on artificial neuronal networks (ANNs) that are in effect just graphs into which input values are fed and combined using simple mathematical operations (often as simple as summation) at nodes, with values being passed from one node to the next (e.g., from one 'layer' to the next) via the graph edges. This captured is captured in popular machine learning libraries, such as Tensor Flow, that allow efficient calculation to be performed across large scale graphs. This graph-based representation extends to more than just ANNs and even simple mathematical functions can be considered as a graph, as in the Figure below.
>
> 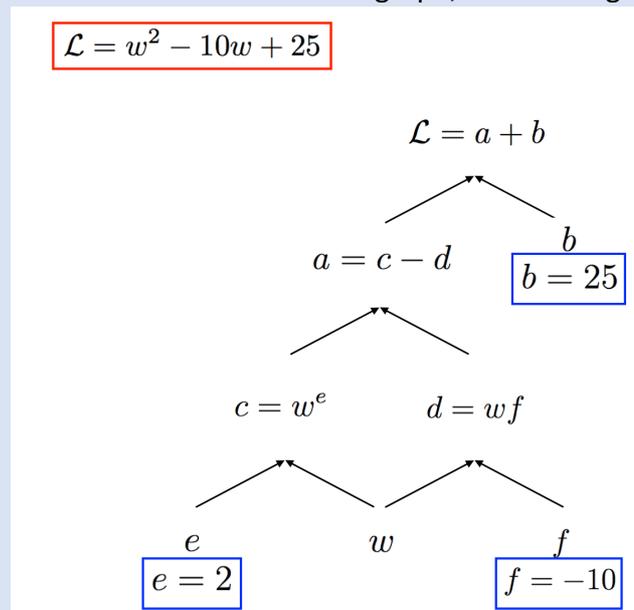
>
> *With thanks to Oiwi for this example*
>
> If we supply a value for *w*, this graph will calculate *L* according the formula. For the purposes of this tutorial the inputs to our function/graph are the data values that once they have passed through the graph produce a calculation of the likelihood.
> By breaking down the function into its 'component' parts we can then exploit AD or backpropagation to calculate the gradient of the function. In essence the gradient calculation involves making a backward pass through the network combining (via the chain rule) the individual differentials of each of the component sub functions represented by each node.

We now have the makings of a general, plausible and potentially efficient scheme for variational inference. However, we are still reliant on a stochastic process and thus 'noisy' gradient estimates that could lead to poor or even failed convergence. Clearly one way to reduce the variability would be to choose *L* to be large. However, this would require a large number of computations at each gradient descent step. Thus, we should make some other attempts at reducing the variability in the gradients we calculate. One thing we can use is the 'reparameterization trick' (Kingma and Welling 2013) the resulting gradient estimates being called by some 'reparameterization gradients'.

The reparameterization trick is something we use to randomly draw $\theta^{*l}$. It allows us to deterministically generate a sample $\theta^{*l}$ from an independent random parameter $\epsilon$. This means that the stochastic process, i.e., draws from $\epsilon$, do not depend upon the hyper-parameters that need to be estimated. This has the effect of the reducing the variability in the estimated gradients when we calculate them using, for example, back-propagation (see Box 4).

> **Box 4: Why does reparametization reduce variance in gradient estimates?**
>
> The fundamental problem with our strategy is that we compute the gradient with respect to $\zeta$ based on *L* samples drawn from the approximate posterior distribution. For this approximation to hold we need a selection of samples that together provide a good approximation. However, the distribution from which we are drawing our samples is itself dependent upon our current estimate of $\zeta$, say from $q^0(\theta)$. If that estimate is poor, e.g., a bad choice of initial value, the majority of our samples might come from values of $\theta$ that even taken together do not produce a truly representative approximation for the gradient, because the 'true' distribution for $q(\theta)$ actually has most of its probability mass elsewhere in $\theta$ space. To get a good estimate of the gradient we would be highly reliant on the rare samples from the tails of the 'poor' distribution $q_\zeta^0(\theta)$ that happen to overlap with the middle of the true distribution. Strictly, our approximation in equations (5.) and (6.) are only an unbiased estimate if we draw $\theta$ from the true $q(\theta)$, but we cannot do that because this is precisely the distirbution we are trying to find. If we start with a bad approximation there is no guarantee that we will get it right in the end (expect in the limit that *L* tends to infinity).
>
> The reparametrisation trick seeks to address this problem. We rewrite $\theta$ as a function of a random variable $\epsilon$ with a distribution, *p*, that does not depend on $\zeta$. Now we can rewrite the expectation (see Box 1) as one over *p* rather than $q(\theta)$. The trick works well when we can choose *p* such that getting a good estimate does not depend on drawing rare values of $\epsilon$. This is facilitated (but not guaranteed) by the fact that *p* does not depend on $\zeta$ and that we can choose *p* to be a simple unimodal distribution. But, there are other cases where it also may work well including where there might be values of $\epsilon$ that are 'important' for a good gradient estimate, but these values are not seen in our generative model, thus they are not 'important' in practice and thus not important in the optimization process.
>
> A fairly general way to find a suitable reparameterization is to exploit the probability integral transform, which tells us that samples a random variable from any arbitrary probability density function will be unfirmly distributed under transformation to the cumulative density function (CDF). Thus, we can get a sampled value for $\theta$ from a sampled value $\epsilon$ via a transformation of the form
>
> $$\theta^* = F_q^{-1}\big[F_p(\epsilon)\big]$$
>
> Where $F_p(\epsilon)$ is the CDF for the distribution *p*, and $F_q(\theta)$ for *q*. This transformation relies on being able to form the inverse of the CDF for q, which conveniently can be done analytically for a normal (and multivariate) normal distribution. It is quite common to choose p to be the standard normal, i.e., $\epsilon \sim N(0,1)$, since we can conveniently sample from this distribution.
>
> Some (if not many) people appear to argue that the reparametrization trick is *necessary* to be able to do back-propagation across the network that forms our objective function when, as in this case, the function includes a stochastic node that itself depends upon deterministic nodes. In this case the stochastic node being the approximate posterior from which we are drawing samples that in turn depends upon hyper-parameters that we want to calculate the gradient with respect to. Thus, we need back-propagation to traverse the stochastic node. The reparamertization trick 'moves' the stochastic part of the process into a separate node, so that the mode representing the approximate posterior is now deterministic and thus the back-propagation can traverse it. However, whilst the reparametrization trick does indeed alter the network in this way, it doesn't appear that this is necessary to allow back-propagation, merely that it results in less variable gradient estimates.

For example, if we were to choose as our approximate posterior a normal distribution with hyper-parameters as the mean, *m*, and the standard deviation, *s*:

$$q(\theta) \sim N(\theta; m, s^2) \tag{7.}$$

Then the reparameterization trick allows us to generate a sample $\theta^{*l}$ from this distribution using:

$$\theta^{*l} = m + s\epsilon \qquad (8.)$$

With $\epsilon \sim N(0,1)$. In principle this 'trick' can be applied to other distributions as long as it is possible to separate the stochastic component from the distribution hyper-parameters. For example, Box 5 extends this to the Multi-Variate Normal distribution, other distributions are also possible see (Ruiz et al. 2016).

We are left with a choice for *L*, the number of sample we will use to calculate our estimated gradient. We expect larger *L* to give a more accurate estimate, but smaller *L* will result in faster computation. In practice, using a simple sample to estimate the free energy gradient (i.e. *L* = 1) may be sufficient; at least in part because we do our optimization over a series of iterations (aka training epochs in Machine Learning parlance), so we can potentially cope with imperfect gradient estimations, i.e. the stochastic nature of using low *L*, over the course of the convergence process. We can assist this process by choosing a variant on Gradient Descent that is particularly designed for stochastic optimization such as the Adam algorithm (Kingma and Welling 2013).

A further thing we can do to potentially improve computational efficiency, but also to circumvent issues with small *L*, is to use 'mini-batches'. This involves dividing the data into subsets and performing a step of the optimization on each batch in turn. Under this method we pass through the data taking multiple steps, then take another pass through the data on a subsequent epoch, again processing one batch at a time. This is a very common approach in Machine Learning problems where the data is 'large' and thus the computation of the cost function (and therefore the gradients too) will be computationally expensive. It can be favourable to only consider a subset at a time, even if this results in more iterative steps

> **Box 5: Approximating the posterior using a multivariate Normal distribution**
> A more useful case than using a univariate normal distribution as the approximate posterior using a multi-variate Normal distribution for approximation to a multi-parameter posterior:
> 
> $$q(\boldsymbol{\theta}) \sim MVN(\boldsymbol{m}, \boldsymbol{C})$$
> 
> Where for a *P* parameter inference, **m** is a (*P* x 1) matrix of parameter means, and **C** is a (*P* x *P*) matrix of the parameter's covariance. Since the covariance matrix should be positive definite we can reparameterise it in terms of a Cholesky decomposition:
> 
> $$\boldsymbol{C} = \boldsymbol{S}\boldsymbol{S}^{\mathrm{T}}$$
> 
> With **S** a (*P* x *P*) lower triangular matrix with positive diagonal entries. In order to ensure the diagonal entries are positive, we parameterise those in log-space:
> 
> $$S(i,i) = e^{v_i}$$
> $$S(i,j) = u_{ij} \text{ for } i > j$$
> 
> In summary, the overall set of hyper-parameters, $\zeta$, that we have to describe the approximate posterior distribution are:
> 
> $$m_i \text{ for } i = \{1 \dots P\}$$
> $$v_i \text{ for } i = \{1 \dots P\}$$
> $$u_{ij} \text{ for } i > j \text{ and } i,j = \{1 \dots P\}$$
> 
> For the MVN we can implement the reparameterisation trick by drawing samples of $\theta$ from
> 
> $$\theta^{*l} = \boldsymbol{m} + \boldsymbol{S}\boldsymbol{\epsilon}$$
> 
> Where $\epsilon \sim MVN(\boldsymbol{0}, \boldsymbol{I})$.

toward convergence. But, it may well be that only using subsets of the data is good enough to achieve convergence in a reasonable number of iterations but will lower computational cost overall.

One of the powerful aspects of sVB is that it naturally exploits a graph-based representation of functions, and thus is amenable to backpropagation, making it compatible with popular and increasingly efficient ML libraries. In fact, the sVB method outlined here is in effect the same as the Variational Auto Encoder, now widely used in ML, see Box 6.

> **Box 6: Why is sVB like the Variational Auto Encoder**
> The main place you might otherwise meet the sVB method considered in this tutorial is in the realm of machine learning, where the same concepts are applied to create what is called the Variational Auto Encoder that is popular for image processing applications. An Auto Encoder is the combination of an Encoder and a Decoder: the Encoder takes data (e.g. an image) as its input and passes it through a network/graph to produce a reduced representation (for images the network would typically involve many layers of convolution and pooling), the Decoder reverses that process and given the representation produces a complete set of predicted data via another (decoding) network. Under ideal conditions the Autoencoder when provided with data should produce at its output an identical set of values that match the data, and the training of the variables in the network can proceed on the basis of minimising a loss function defined on the difference between true and predicted data.
> The Variational Auto Encoder adds a constraint on the encoding network that forces it to generate representations that follow a probability distribution (the original version made this a Gaussian distribution). This allows it to generalise from simply memorising the inputs it has been shown during the optimization (training) stage, to being able to generate new predictions having 'learned' something about the data it has been provided.
> This is analogous to the sVB method where our approximate posterior distribution is performing the function of the constrained representation, our model is the encoder and the Free Energy is the loss function that we evaluate in the optimization of the parameters of our representation. By drawing from the posterior, as we do to calculate the loss function, we in effect do decoding to produce a prediction. Typically, the VAE, which is composed of two ANNs, has many variables associated with edges in the encoding and decoding graphs that need to be optimised. For the sVB formulation, the equivalent' networks are fixed, apart from a small number of hyper-parameters that we estimate.

# Example 1 – fitting a Gaussian distribution

First, we consider the simple (and classic) case of inferring on a single univariate Gaussian distribution from some data. We will attempt to infer the (approximate) joint full posterior of both the mean and variance of the Gaussian from which are data is drawn.

> Note, this case is exactly analogous to the familiar data analysis scenario of having a number of noisy measurements of a single quantity and wanting an estimate of both the mean and variance of the measurements, i.e., a 'best estimate' of the quantity being measured and also a measure of the noise magnitude.

## Generative Model

Our measurements come from a Gaussian distribution with mean, $\mu$, and precision (1/variance), $\beta$:

$$P(y_n|\mu,\beta) = \frac{\sqrt{\beta}}{\sqrt{2\pi}} e^{-\frac{\beta}{2}(y_n-\mu)^2}$$

If we draw *N* samples that are identically independently distributed (i.i.d) we have:

$$p(\mathbf{y}|\mu,\beta) = \prod_{n=1}^{N} p(y_n|\mu,\beta) = \left(\frac{\beta}{2\pi}\right)^{\frac{N}{2}} e^{-\frac{\beta}{2}\sum_{n=1}^{N}(y_n-\mu)^2}$$

## Priors

Unlike 'traditional' VB we are not restricted to conjugate priors. In this example, we somewhat arbitrarily choose a MVN prior over the mean, $\mu$, and the log of the variance, $\log\left(\frac{1}{\beta}\right)$:

$$\begin{bmatrix}\mu \\ -\log(\beta)\end{bmatrix} \sim MVN(\mathbf{m}_0, \mathbf{C}_0)$$

where we will choose the prior to be fairly noninformative by selecting hyper-parameters:

$$\mathbf{m}_0 = \begin{bmatrix}0 \\ 0\end{bmatrix}, \quad \mathbf{C}_0 = \begin{bmatrix}100 & 0 \\ 0 & 100\end{bmatrix}$$

> 'Traditional' VB would call for a normal distribution for $\mu$, and a gamma distribution for $\beta$ and thus require a 'mean field' approximation: a posterior distribution made up of a product of two independent distributions. Resulting in no possibility of inferring correlation between the parameters).

## Approximate Posterior

Again, there is no restriction to conjugate distributions, thus we are also free to choose our approximating posterior. A MVN will be convenient, partly because we know how to interpret the hyper-parameters of an MVN, but also because we know the reparameterization trick will be possible. As defined by our choice above, the parameters in $\theta$ are $\mu$ and $\log(1/\beta)$, hence:

$$q(\theta) = q\begin{pmatrix}\mu \\ -\log(\beta)\end{pmatrix} \sim MVN(\mathbf{m}, \mathbf{C})$$

Where **m** is a (2x1) matrix of estimated means of the parameters of the Gaussian distribution and **C** is a (2x2) matrix containing the estimated covariance matrix for the parameters of the Gaussian distribution, i.e. it tells us about the uncertainty with which we can estimate both the mean and precision of the Gaussian distribution that is generating the data.

**Free energy**

We now have all of the information we need to write down the terms in the Free energy in equation (3.) and thus implement the approximation in (and in turn the gradient calculations of) equation (5.). The log-likelihood is:

$$\log(p(y|\boldsymbol{\theta})) = \frac{N}{2}\log\frac{\beta}{2\pi} - \frac{\beta}{2}\sum_{n=1}^{N}(y_n - \mu)^2$$

And the log KL-divergence between (approximate) posterior and prior:

$$\log\left(\frac{q(\boldsymbol{\theta})}{p(\boldsymbol{\theta})}\right) = -\frac{1}{2}\log\left(\frac{|\boldsymbol{C}|}{|\boldsymbol{C}_0|}\right) - \frac{1}{2}(\boldsymbol{\theta} - \boldsymbol{m})^{\mathrm{T}}\boldsymbol{C}^{-1}(\boldsymbol{\theta} - \boldsymbol{m}) - \frac{1}{2}(\boldsymbol{\theta} - \boldsymbol{m}_0)^{\mathrm{T}}\boldsymbol{C}_0^{-1}(\boldsymbol{\theta} - \boldsymbol{m}_0)$$

By choosing a MVN for both the prior and the approximate posterior distributions we can perform the required integral of this second part of equation (5.) (i.e., we can compute analytically the expectation of the log KL-divergence with respect to the approximate posterior, See Box 7):

$$Loss_L = \int q(\boldsymbol{\theta})\log\left(\frac{q(\boldsymbol{\theta})}{p(\boldsymbol{\theta})}\right)d\boldsymbol{\theta}$$

$$= \frac{1}{2}\left\{\mathrm{Trace}(\boldsymbol{C}_0^{-1}\boldsymbol{C}) - \log\left(\frac{|\boldsymbol{C}|}{|\boldsymbol{C}_0|}\right) - N + (\boldsymbol{m} - \boldsymbol{m}_0)^{\mathrm{T}}\boldsymbol{C}_0^{-1}(\boldsymbol{m} - \boldsymbol{m}_0)\right\}$$

This means we do not need a stochastic approximation to this part of the Free energy, i.e. we can compute this expression in place of equation (5.):

$$F \approx Loss_L + \frac{1}{L}\sum_{L}\log(p(y|\theta^{*l}))$$

Which we might hope will make our approximations more accurate and less variable.

> Note that, this step is not a result of the trivial nature of the problem we are considering in this example, as it doesn't depend on the likelihood (and thus generative model) at all. It arises from the choice of prior and posterior distributions, which will thus generalise to other implementations.

## Using mini-batches

We can setup the sVB inference so that it proceeds via mini-batches of the data in an attempt to arrive at a faster solution. In doing this we have to take care, since now we will pass only a subset of the data to the function that calculates the log-likelihood. If we do not rescale the resulting value it will now be smaller in proportion to the other term in the Free Energy arising from the KL-divergence that includes the prior, and thus the prior will have greater weight in the final estimated posterior akin to only doing inference on a smaller data set. Thus, if we have a batch size of *M* (where *M* < *N*) the log-likelihood is:

$$\log(p(y|\boldsymbol{\theta})) = \frac{N}{2}\log\frac{\beta}{2} - \left(\frac{N}{M}\right)\frac{\beta}{2}\sum_{m=1}^{M}(y_m - \mu)^2$$

> **Box 7: Finding the expectation of the log-KL divergence term in the Free energy**
>
> $$Loss_L = \int q(\boldsymbol{\theta})\log\left(\frac{q(\boldsymbol{\theta})}{p(\boldsymbol{\theta})}\right) d\boldsymbol{\theta}$$
>
> $$= -\frac{1}{2}\int \left\{\log\left(\frac{|C|}{|C_0|}\right) + (\boldsymbol{\theta}-\boldsymbol{m})^T C^{-1}(\boldsymbol{\theta}-\boldsymbol{m}) - (\boldsymbol{\theta}-\boldsymbol{m}_0)^T C_0^{-1}(\boldsymbol{\theta}-\boldsymbol{m}_0)\right\} q(\boldsymbol{\theta}) d\boldsymbol{\theta}$$
>
> $$= -\frac{1}{2}\left\{\log\left(\frac{|C|}{|C_0|}\right) + Trace(C^{-1}C) - \int [(\boldsymbol{\theta}-\boldsymbol{m})^T C_0^{-1}(\boldsymbol{\theta}-\boldsymbol{m}) - (\boldsymbol{\theta}-\boldsymbol{m})^T C_0^{-1}\boldsymbol{m}_0 - \boldsymbol{m}_0^T C_0^{-1}(\boldsymbol{\theta}-\boldsymbol{m}) + (\boldsymbol{m}-\boldsymbol{m}_0)^T C_0^{-1}(\boldsymbol{m}-\boldsymbol{m}_0)] \; q(\boldsymbol{\theta}) d\boldsymbol{\theta}\right\} =$$
>
> $$= -\frac{1}{2}\left\{\log\left(\frac{|C|}{|C_0|}\right) + N - \text{Trace}(C_0^{-1}C) - (\boldsymbol{m}-\boldsymbol{m}_0)^T C_0^{-1}(\boldsymbol{m}-\boldsymbol{m}_0)\right\}$$
>
> Where we have used the following general results:
>
> $$\int (x-m)^T U^{-1}(x-m)\, MVN(x;m,V) dx = Trace(VU^{-1})$$
>
> $$\int (x-m) MVN(x;m,V) dx = m - m = 0$$

## Methods

We implemented the posterior and associated reparameterisation trick using the approach in Box 5. This means that we will have a total of 5 hyper-parameters: two for *m* representing the two entries in *m*, and 3 for *C* arising from the decomposition meaning we only need two diagonal values and one off-diagonal. We also a consider a variant of this with only two parameters for *C*, i.e., non-zero only on the diagonal, thus not estimating any correlation between the model parameters in the posterior.

The stochastic VB inference method was implemented in python using the Tensor Flow library (v1.4), using the Adam optimizer. Data were generated in the form of 100 samples from the likelihood with (mean) $\mu = 1$ and (variance) $1/\beta = 4$. Optimization was run for 400 epochs and two different strategies were considered: full data inference, where on each epoch a single optimisation step is taken using the full data (i.e. 400 iterations are performed); mini-batches, the data is divided into 10 batches of 10 data points and optimisation is performed on each sequentially, thus on each epoch was pass through the data once by performing 10 optimization steps (4000 iterations in total).

## Results

Figure 1 shows the results of from the inference on a set of sampled data (as shown in Figure 1(a)) comparing the estimate posterior from 2D grid search (b, for this no prior was defined so effectively this is just the likelihood), stochastic VB with (d) and without (c) inference of the correlation of the two parameters in the posterior. All three methods correctly identify that the mean the mean and variance of the data. Noticeably, the form of the estimated posterior from the stochastic VB inference, which is a parameterised MVN distribution, matches very closely the sampled distribution from the 2D grid search. Figure 2 shows the approximate Free Energy, the cost function, evaluated at each epoch. For both sVB inferences convergence appears to be reached somewhere between epoch 50 and 100. Note, though, that it is hard to judge convergence because the Free energy values are themselves stochastic estimates based on only a single sample of $\theta$, hence they quite variable. Further implications of this are that we cannot select the epoch with the lowest Free energy, since that might not represent the best solution, but could simply be the result of the sample evaluated in that epoch. Additionally, we might need to take care if we want the value of free energy for any further calculations, for example in using it as an approximation to the model evidence for model comparisons. In that scenario we should calculate the free energy using a larger number of samples.

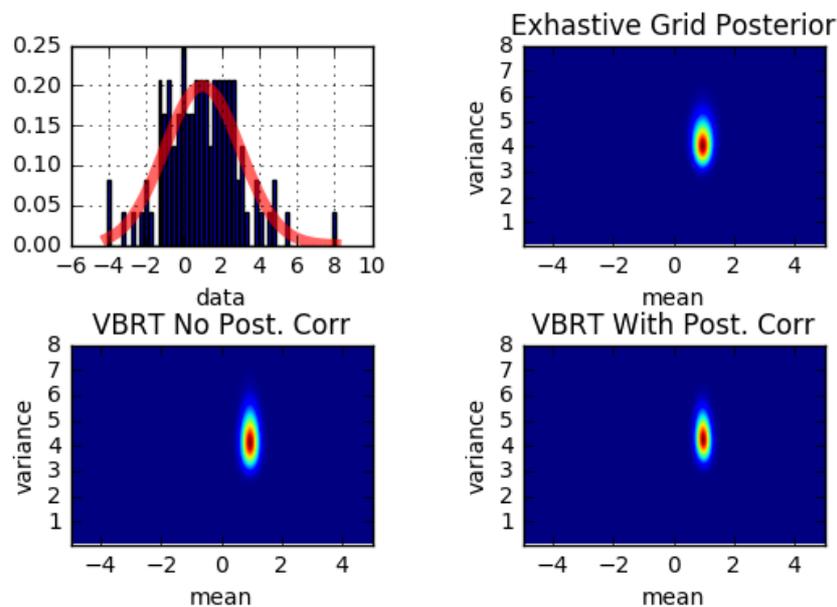

*Figure 1: Results from Bayesian inference on data generated from a normal distribution. (a) true distribution (red) and histogram of sampled data (blue), (b) posterior distribution on mean and variance evaluated using 2D grid evaluation, (c &d) posterior distribution estimated using sVB without and with a hyper-parameter to capture correlation between the two parameters.*

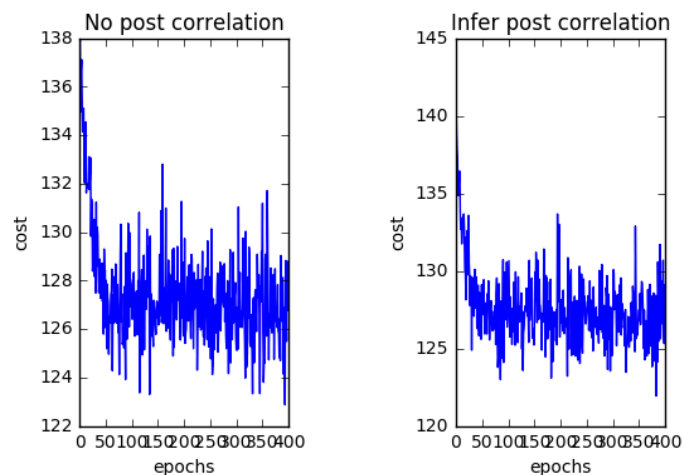

*Figure 2: (approximate) Free Energy calculated at each epoch for both variants on sVB inference.*

Figure 3 shows the results when using mini-batches. The results are strikingly similar to the case without mini-batches, with perhaps the only difference being in the case where a correlation of the parameters was allowed for in the posterior. Figure 4 shows the Free Energy at each epoch. It appears that convergence in this case happens in far fewer epochs, quite possibly within the first 10 epochs (subject to the caveats mentioned above). It is worth remembering, however, that each epoch now involves 10 separate optimisation steps. Thus, the overall number of calls to the Free energy calculation (and gradient calculation) may not be that different between the two cases. But, each call in this mini-batch case does only involved one 10$^{th}$ of the data, so might be expected to be faster. Thus, even in this trivial example the use of mini-batches appears to be advantageous. (But, trying to measure this on such a small problem is probably futile).

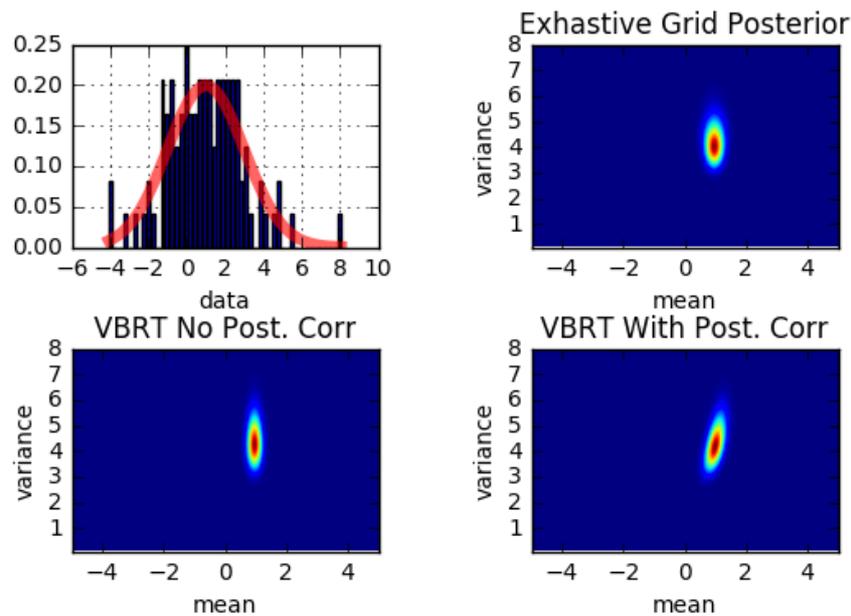

Figure 3: Results from Bayesian inference on data generated from a normal distribution. (a) true distribution (red) and histogram of sampled data (blue), (b) posterior distribution on mean and variance evaluated using 2D grid evaluation, (c &d) posterior distribution estimated using sVB employing mini-batches, without and with a hyper-parameter to capture correlation between the two parameters.

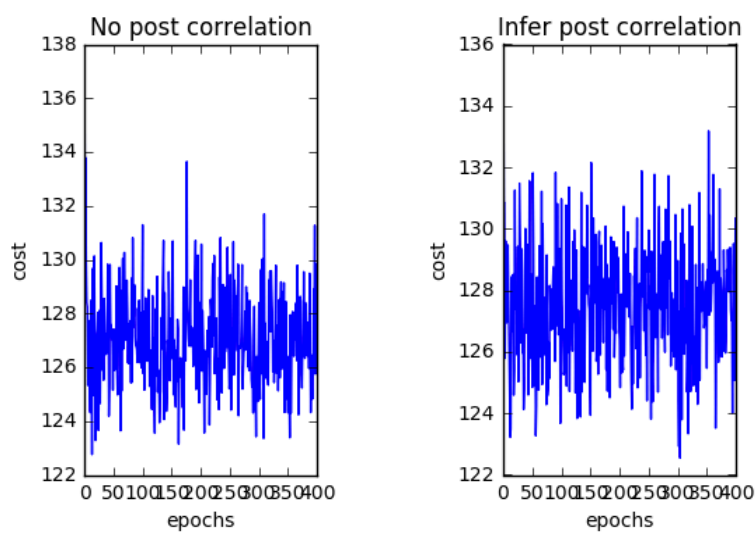

Figure 4: (approximate) Free Energy calculated at each epoch for both variants on sVB inference when employing mini-batches (note, each epoch involves 10 iterations of the optimization algorithm).

## Example 2 – Inferring a Folded Normal distribution

For the second example, we consider data generated from a Folded Normal distribution: a distribution where only positive values are possible. One reason for choosing this example is that for data drawn from this distribution we would expect correlation between the mean and variance of the distribution in our posterior. The other reason is that we would not be able to apply 'traditional' VB to this example because we would not be able to find a distribution that gives conjugacy between likelihood and prior.

### Generative model

We draw measurement from a Folded Normal distribution with mean, $\mu$, and precision, $\beta$:

$$p(y_n|\mu,\beta) = \frac{\sqrt{\beta}}{\sqrt{2\pi}} e^{-\frac{\beta}{2}(y_n-\mu)^2} + \frac{\sqrt{\beta}}{\sqrt{2\pi}} e^{-\frac{\beta}{2}(y_n+\mu)^2}$$

For $y_n > 0$, and 0 otherwise. An alternative formulation is:

$$p(y_n|\mu,\beta) = \frac{\sqrt{2\beta}}{\sqrt{\pi}} e^{-\frac{\beta}{2}(y_n+\mu)^2} \cosh(\mu y_n \beta)$$

### Prior and Posterior

For this example, we will use precisely the same prior and (approximate) posterior distributions as we did in the first example. Noting that our choice of posterior might not be the best we could use (there may be one that makes a better approximation), but that we are free to choose what we like. Since it is only the likelihood that has changed, we can reuse the result derived in the first example for the KL-divergence between posterior and prior.

### Results

Figure 5 shows a set of results from inference of data drawn from a folded normal distribution. For this dataset the sVB method estimates a slightly higher mean and lower variance than was used to generate the data (this is reasonably variable depending upon the data used). The Inference with posterior correlation captures more of the correlation between the mean and variance parameters, i.e., reflecting that there are multiple values of mean and variance of the distribution that could plausibly explain the data. There is a noticeable discrepancy between the sVB solution and the grid search in this case, which might partially be explained by the lack of influence of a prior on the grid search solution

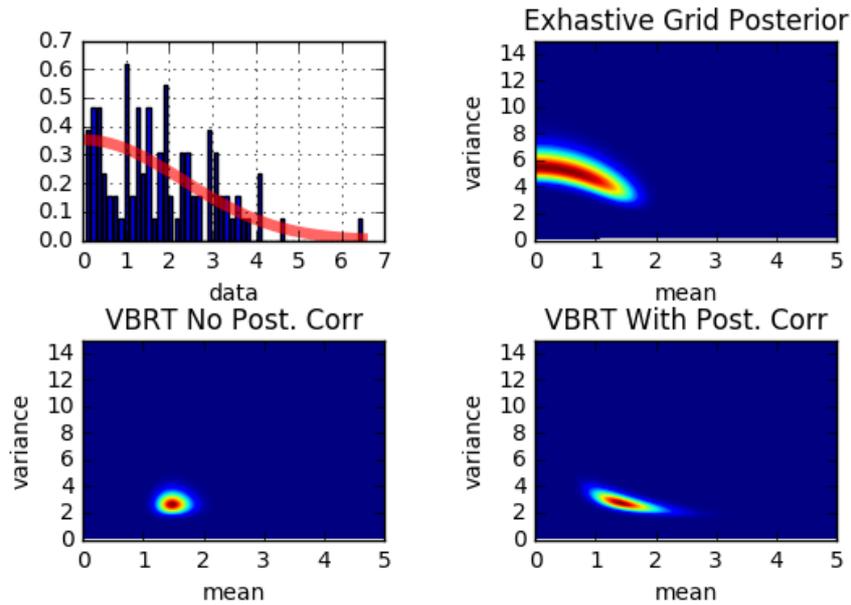

*Figure 5: Results from Bayesian inference on data generated from a Folded Normal distribution. (a) true distribution (red) and histogram of sampled data (blue), (b) posterior distribution on mean and variance evaluated using 2D grid evaluation, (c &d) posterior distribution estimated using sVB without and with a hyper-parameter to capture correlation between the two parameters.*

Figure 6 shows the results when using mini-batches. As before the sVB solution without correlation estimation looks similar to the non-batched analysis, but there are some differences when correlation is included in the inference, in this case it is doing a better job of estimating the mean and variance of the data. Some very empirical exploration suggests that the estimated posterior from the approximated posterior without correlation is stable over different runs (with the same data), but more variability is observed between independent runs when posterior correlation is included. This doesn't appear to be reflected in the values of Free Energy to which the method 'converges'. As in the first example, the apparent convergence of the Free Energy is more rapid with epoch (not shown)

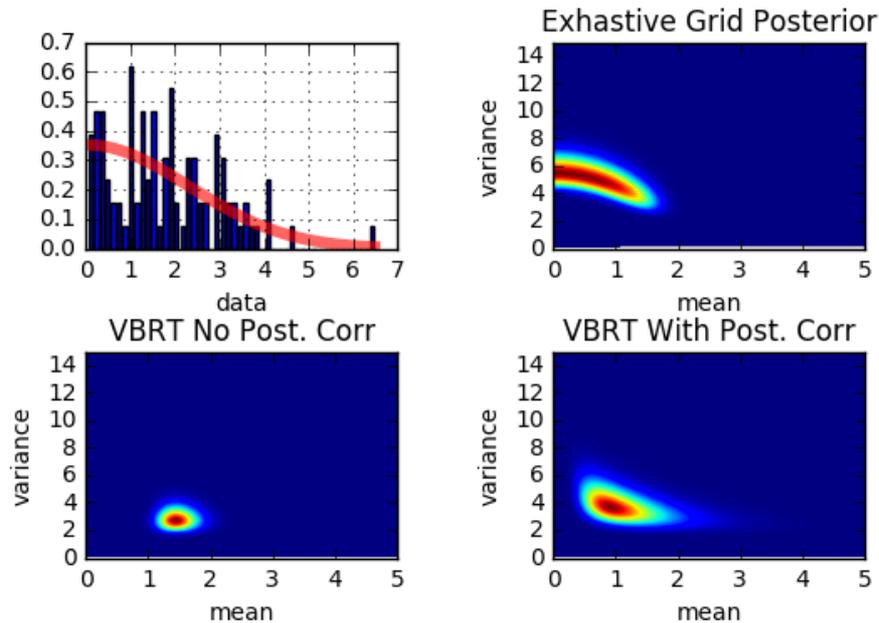

*Figure 6: Results from Bayesian inference on data generated from a Folded Normal distribution. (a) true distribution (red) and histogram of sampled data (blue), (b) posterior distribution on mean and variance evaluated using 2D grid evaluation, (c &d) posterior distribution estimated using sVB employing mini-batches, without and with a hyper-parameter to capture correlation between the two parameters.*

# Conclusions

In this tutorial we have 'updated' our previous introduction to Variational Bayes to a more recent and potentially more flexible approach based on stochastic approximations: stochastic Variational Bayes (sVB). In doing so we can relax the restrictions of the mean-field approximation and conjugacy, and exploit computational advances that are reaping rewards for machine learning methods. We have illustrated the methodology on some simple, and relatively familiar, cases to provide some insight as to how sVB might be deployed for Bayesian Inference.

# References

Attias H, Leen T, Muller K-L. **A variational Bayesian framework for graphical models**. In Advances in Neural Information Processing Systems, vol. 12 p. 49–52. 2000.

Chappell MA, Woolrich MW. **The FMRIB Variational Bayes Tutorial: Variational Bayesian Inference for a Non-Linear Forward Model**, v1.1. Oxford University Research Archive. 2016. https://ora.ox.ac.uk/objects/uuid:8a90a2a5-4748-4557-a6f2-4eee5f8b07ae

Kingma DP, Welling M. **Auto-Encoding Variational Bayes**. arXiv 1312.6114. 2013;

Ruiz FJR, Titsias MK, Blei DM. **The Generalized Reparameterization Gradient**. arXiv 1610.02287. 2016;